\documentclass[apj]{emulateapj}
\usepackage{natbib}
\usepackage{geometry}
\usepackage{amsmath}

\usepackage{graphicx}
\usepackage{ulem,color}

\begin{document}

\title{Testing the magnetar model via late time radio observations
  of two macronova candidates}


\author{
Assaf Horesh\altaffilmark{1}, 
Kenta Hotokezaka\altaffilmark{2},
Tsvi Piran\altaffilmark{2},
Ehud Nakar\altaffilmark{3},
Paul Hancock\altaffilmark{4,5}
}

\altaffiltext{1}{Benoziyo Center for Astrophysics, Weizmann Institute of Science,
76100 Rehovot, Israel}
\altaffiltext{2}{Racah Institute of Physics, The Hebrew University, Jerusalem 91904, Israel}
\altaffiltext{3}{Raymond and Beverly Sackler School of Physics \& Astronomy, Tel Aviv University, Tel
Aviv 69978, Israel}
\altaffiltext{4}{International Centre for Radio Astronomy Research
  (ICRAR), Curtin University, GPO Box U1987, Perth WA 6845, Australia}
\altaffiltext{5}{ARC Centre of Excellence for All-Sky Astrophysics (CAASTRO)}

\begin{abstract}

Compact binary mergers may have
already been observed as they are the leading model for 
short gamma-ray bursts (sGRBs).  Radioactive decay within the ejecta
from these mergers is expected to produce an infra-red flare, dubbed macronova (or
kilonova), on a time scale of a week. Recently two such macronova candidates
were identified in followup observations of sGRBs,
strengthening the possibility that those indeed arise from mergers. 
The same ejecta will also produce a long term
(months to years) radio emission due to its interaction  with 
the surrounding ISM. In search
for this emission, we observed the two macronova candidates,
GRB\,130603B and GRB\,060614 with the Jansky very large array
(VLA) and the Australia Telescope Compact Array (ATCA). Our
observations resulted in null-detections,  putting strong upper  limits
on the kinetic energy and mass of the ejecta. 
A possible outcome of a merger is a 
highly magnetized neutron star (a magnetar), which has been  suggested as the
central engine for GRBs. Such a magnetar 
 will deposit a significant fraction of its energy
into the ejecta leading to a brighter radio flare. Our results,
therefore, rule out magnetars
in these two events.

\end{abstract}

\section{Introduction}
\label{sec:intro}

The coalescence of two compact objects such as a Neutron Star (ns) -
Black Hole (BH) merger or a ns$^{2}$ merger has been a leading
candidate for the progenitor
system for short-duration ($< 2$\,s)
Gamma-ray bursts (sGRBs; Eichler et al. 1989;
Narayan, Paczynski \& Piran 1992; see reviews by Nakar 2007, Berger
2014). 
Li \& Paczynski (1998) suggested that mergers will be accompanied by the so called
``macronova'' (or ``kilonova'').  They suggested that the radioactive decay of the neutron rich matter ejected in a merger
event would lead to a brief ($\sim 1$\,day) optical signal that might
be detectable. More recently, Barnes \& Kasen (2013) and  Tanaka \& Hotokezaka
(2013)  have
revised  the original prediction of Li \& Paczynski. As the optical
depth for r-process elements is high, the optical signal is expected
to be mostly absorbed, and a longer ($\sim 1$\,week)  IR signal is expected  instead.

A second prediction of the merger scenario is of late-time radio
emission (Nakar \& Piran 2011). The same ejecta that produce the macronova is expected to
interact with the  interstellar medium (ISM). The resulting shockwave,
ploughing through the ISM will accelerate electrons and produce
magnetic fields. In turn, this will lead to synchrotron radio
emission, similar to the process responsible for radio emission
from supernovae (e.g., Chevalier 1982; Chevalier \& Fransson 2006) or
GRB afterglows (Sari 1997; Sari et al. 1999). The rise time and the peak flux of this 
radio flare, depend mainly on the ejecta mass, its velocity and on the
density of the ISM. For a macronova ejecta
mass of ${\rm M}_{ej}\sim 0.01\,{\rm M}_{\odot}$ and velocity range of
$v_{ej}\sim 0.1 - 0.3$\,c, as predicted by recent numerical
simulations (Rosswog et al. 2013; Hotokezaka et al. 2013, Bauswein et
al. 2013), the radio emission is expected to rise over a time-scale of
months to years.

A variant of the simple merger scenario, motivated by a strong amplification of magnetic fields at
the merger shown in numerical simulations  (Price \& Rosswog 2006,
Rezzolla 2012, Giacomazzo \& Perna 2013, Giacomazzo et al. 2015, Kiuchi et al. 2015),
 is the magnetar scenario, in which the
remnant of the merger is a highly magnetized ns with a period of $\sim
1$\,ms and a large magnetic field of $B > 10^{13}$\,G (Usov 1992; Rosswog 
\& Davies 2002; Bucciantini et al. 2012; Fan et
al. 2013; Giacomazzo et al. 2013; Metzger \& Piro 2014; Siegel et
al. 2015).  The magnetar has been suggested (e.g., Usov 1992; Duncan
\& Thompson 1992; Zhang \& M{\'e}sz{\'a}ros 2001; Metzger 2010) as the central engine 
that powers GRBs. 
In this scenario, the magnetar deposits most of its 
rotational energy, as it spins down, in the macronova ejecta, accelerating it to high
relativistic velocities, thus significantly enhancing the expected radio
flare signal.  A magnetar with a $\sim 1$\,ms period deposits a 
kinetic energy of  $\approx 3\times 10^{52}$
erg. The time-scale and the peak flux of the expected radio emission
in this case, can be calculated (in a simplistic way; see \S 3) according to the formalism of Nakar \& Piran (2011), but with a high relativistic
velocity ($\beta \equiv v_{ej}/c \approx 1$).  The magnetar radio emission, that will also rise over a long time
scale, is expected to be brighter by a few orders of magnitude than
the non-magnetar merger scenario (as first discussed by Metzger \&
Bower 2013; see \S\ref{sec:sumary}). In addition to the bright radio
emission from the forward shock, there may be additional radio
emission from the pulsar wind nebula, which is expected to peak over
much shorter time scales (see Piro \& Kulkarni 2013; Metzger \& Piro
2014).

Berger et al. (2013) and Tanvir et al. (2013) have identified 
 a macronova candidate associated with GRB\,130603B. The macronova signal
was detected, as predicted, in the IR band and  it lasted less than 30
days\footnote{HST observations detected the IR source 9 days after the
  sGRB was detected. A second observing HST epoch undertaken 21
  days later revealed that the source have faded away (Tanvir et
  al. 2013)}. Berger et al. (2013) find that an ejecta with a mass of
${\rm M_{ej}} = 0.05 - 0.08\,{\rm M}_{\odot}$ and velocities of
$v_{ej} \approx 0.1 - 0.3$\,c is needed\footnote{Note that a wider range
  of ejecta mass and velocities is consistent with the data.} to produce the observed signal. A second macronova candidate
associated with the earlier event GRB\,060614, has been
recently discovered by Yang et al. (2015) who re-examined the data
obtained with the Hubble Space Telescope and found excess
emission at the F814W band (see also Jin et al. 2015). Yang et al. find that it can be
explained by a macronova with a significantly more massive ejecta, ${\rm M_{ej}} = 0.03 - 0.1\,{\rm M}_{\odot}$, with velocities of $v_{ej} \approx 0.1 - 0.2$\,c.

In search for a late-time radio emission, originating from a forward
shock in the ISM, as predicted by Nakar \& Piran
(2011), we obtained late-time
radio observations of both GRB\,130603B and GRB\,060614. In the next
section we briefly describe the radio observations. In \S 3 we provide details about how the predictions of the radio signal are
calculated. We compare our predictions with the results of our
observations in \S 4, and briefly summarize in \S 5.

\section{Radio observations}
\label{sec:obs}

\subsection{VLA observations of GRB\,130603B}
\label{sec:vla}

We observed GRB\,130603B with the Karl G. Jansky Very Large Array
(VLA), in B configuration, on 2015, February 12 UT ($T_0+619$\,days). The
observation was performed at central frequency of $3$\,GHz using J\,1120+1420 and 3C286 as phase and flux calibrators,
respectively. We analysed the data using standard AIPS\footnote{Astronomical Image Processing System}
and CASA\footnote{Common Astronomy Software Applications package; McMullin et al. 2007} routines. We found no significant radio
emission at the position of the GRB with a $3\sigma$ detection limit
of $60\mu$Jy. 

At early times, GRB\,130603B was observed with the VLA by Fong et
al. (2014). They detected radio emission a few hours after the GRB was
discovered. This emission was rapidly fading away below the detection
limit, within four days as expected from a typical GRB afterglow. 
An observation at day $84$ after discovery  (Fong et
al., 2014) 
resulted in a null-detection of $34$\,$\mu$Jy ($3\sigma$) at
$6.7$\,GHz. 

\subsection{ATCA observations of GRB\,060614}
\label{sec:atca}

We used the Australia Telescope Compact Array (ATCA)  to observe
GRB\,060614A at $2.1$\,GHz on 2015 May 9 UT ($T_0+337$\,days). The
following calibrator sources: PKS\,B1921$-$293 (band-pass),
PKS\,B1934$-$638 (flux), and PKS\,B2213$-$45 (phase) were used. The
data were processed using {\sc miriad} (Sault, Teuben \& Wright 1995).
The resulting image achieved an RMS noise of $50$\,$\mu$Jy at the location of the GRB, however no detection was made with a $3\sigma$ upper limit of $150$\,$\mu$Jy.

\section{Prediction of the magnetar signal}
\label{sec:model}

The interaction  of the ejected mass with the ISM leads to a late time (months-years) radio signal. The luminosity
of the signal and the peak time depend both on the properties of the
ejecta and of the ISM. At high frequencies the signal is expected to peak once the ejecta starts to
decelerate. This will occur when the ejecta ploughed through
sufficient 
ISM mass to slow it down, i.e., comparable to the ejecta mass. The
deceleration radius and time, and the peak radio flux are then simply
defined by Nakar \& Piran (2011).

In the magnetar scenario, the stable ns remnant formed in a binary ns merger is expected to have a
typical rotational period of $P \sim 1$\,ms. The rotational energy of
the ns is 
\begin{equation}
E_{\rm rot}=\frac{I (2\pi)^{2}}{2 P^{2}} 
\end{equation}
where $I$ is the moment of inertia. For the above period, the
rotational energy is $\approx 3\times 10^{52}$\,erg. 
Depositing this additional energy into an ejecta mass of $10^{-2}{\rm
  M}_{\odot}$ will result in a relativistic outflow, leading to a
stronger 
radio signal at late times. It is important to stress that our
estimates are sensitive just to the magnetar period and not to its
magnetic field. The total rotational energy of the magnetar is released and deposited in the ejecta on a time scale much shorter than the time scales that we consider here. 

Adopting the above typical magnetar energy and assuming
the ejecta mass and ISM density, the predicted radio light
curves can be calculated using the Nakar \& Piran (2011) formalism (see also Piran et al. 2013). However, there are two additional points that
need to be treated more carefully. First, the peak flux is given
assuming that the observed frequency is above the self-absorbed
frequency. Second, Nakar \& Piran address the case of non-relativistic
ejecta and thus neglected relativistic effects.

The major relativistic effects on the
observed flux are: (i) relativistic time effects, (ii) the Doppler shift,
and (iii) relativistic beaming\footnote{In the case of an isotropic
  ejecta, the relativistic beaming does not change the total luminosity.}~(see e.g. Piran 2004 for a review). 
These effects play important roles depending on the initial Lorentz
factor. Roughly
speaking, an observer will measure  
a brighter flux than those expected from Newtonian motion until the blast wave has sufficiently decelerated.
The deceleration timescale, in the relativistic case, is shorter than
the Newtonian one by a factor of 
$\Gamma^{-8/3}$. 
The synchrotron frequency $\nu_{m}$  corresponding to $\gamma_{m}$\footnote{We have assumed that the electrons are accelerated by the blast-wave  with a power-law energy distribution of $N_{\rm e}\sim\gamma_{\rm
  e}^{-p}$, with some minimum Lorentz factor $\gamma_{m}$.} decreases with time. For an observed  frequency $\nu$ that 
  was initially below $\nu_m$ the flux peaks when $\nu=\nu_m$ at: 
\begin{equation}
\begin{aligned}
& t_{\rm peak}  = ~ 120~{\rm days}\left(\frac{E}{3\cdot 10^{52}} \right)^{1/3}\left(\frac{\epsilon_{e}}{0.1} \right)^{4/3}\left(\frac{\epsilon_{B}}{0.1} \right)^{1/3}\\
& \left(\frac{\nu}{3~{\rm GHz} }\right)^{-2/3}.
\end{aligned}
\end{equation}
where, $E$ is the energy deposited in the ejecta, $\epsilon_{B}$ and
$\epsilon_{e}$ are the shock equipartition parameters of the magnetic
field and electron energy, respectively. The peak flux at this time is 
\begin{equation}
\begin{aligned}
& F_{\nu,{\rm peak}}  = ~ 8~{\rm mJy}\left(\frac{E}{3\cdot 10^{52}} \right)\left(\frac{\epsilon_{B}}{0.1} \right)^{1/2}\left(\frac{n}{0.1~{\rm cm^{-3}}} \right)^{1/2}\\
& \left(\frac{D}{10^{28}~{\rm cm}} \right)^{-2}
\end{aligned}, 
\end{equation}
where $D$ is the distance to the source, and $n$ is the ISM
density. 
The above estimates are valid when synchrotron self-absorption
is negligible. The radio frequencies are often 
below the self-absorption frequency $\nu_{a}$. For $\nu_{m}<\nu <\nu_{a}$
 the peak time (after the deceleration) is when $\nu=\nu_{a}$:
\begin{equation}
\begin{aligned}
& t_{\rm peak}  = ~ 170~{\rm days}\left(\frac{E}{3\cdot 10^{52}} \right)^{\frac{p+2}{3p+2}}\left(\frac{\epsilon_{e}}{0.1} \right)^{\frac{4(p-1)}{3p+2}} \\  
& \left(\frac{n}{0.1~{\rm cm^{-3}}} \right)^{\frac{4}{3p+2}}\left(\frac{\epsilon_{B}}{0.1} \right)^{\frac{p+2}{3p+2}}\left(\frac{\nu}{3~{\rm GHz} }\right)^{\frac{-2(p+4)}{3p+2}},
\end{aligned}
\end{equation}
and the peak flux is estimated as
\begin{equation}
\begin{aligned}
& F_{\nu,{\rm peak}}  = ~ 5~{\rm mJy}\left(\frac{E}{3\cdot 10^{52}} \right)^{\frac{2p+3}{3p+2}}\left(\frac{\epsilon_{e}}{0.1} \right)^{\frac{5(p-1)}{3p+2}} \\
& \left(\frac{n}{0.1~{\rm cm^{-3}}} \right)^{\frac{8-3p}{6p+4}}\left(\frac{\epsilon_{B}}{0.1} \right)^{\frac{p+4}{6p+4}}\left(\frac{\nu}{3~{\rm GHz} }\right)^{\frac{5p-5}{3p+2}}\\
& \left(\frac{D}{10^{28}~{\rm cm}} \right)^{-2}.
\end{aligned}
\end{equation}

In order to account for the blast wave dynamics in both relativistic and
non-relativistic regime, we follow the numerical procedures of
Hotokezaka \& Piran (2015). Note, that we do not use directly the
approximate equations above, but perform a full numerical
calculation. In short (see the following references for
more details), the blast wave expansion is determined by conservation of energy of the ejecta and
swept-up material as $M(R)(\Gamma \beta c)^{2} = E$, in a similar way to Piran et al. (2013).
Here $M(R)$ is the sum of the ejecta mass and the mass of the swept-up
ISM at a radius $R$.  
For a given blast-wave dynamics we calculate the synchrotron radiation, using the fluid velocities and energy density just behind the shock. Then, at each observer time, we sum up the emission from each fluid element following Eq. (4) of Granot, Sari, \& Piran
1999a. This  includes all relativistic propagation effects consistently and reproduces the light
curves of Sari et al. (1998) in the ultra-relativistic limit and of  Nakar \& Piran (2011) in the non-relativistic
limit. To account for synchrotron self-absorption, we calculated the
absorption coefficients based on Granot, Piran, \& Sari (1999b). In
addition, throughout our calculations, we adopt the following parameter values: $\epsilon_{e}=0.1$ 
and $p=2.5$.

\begin{figure*}[!ht]
\centering
\includegraphics[width=0.8\textwidth]{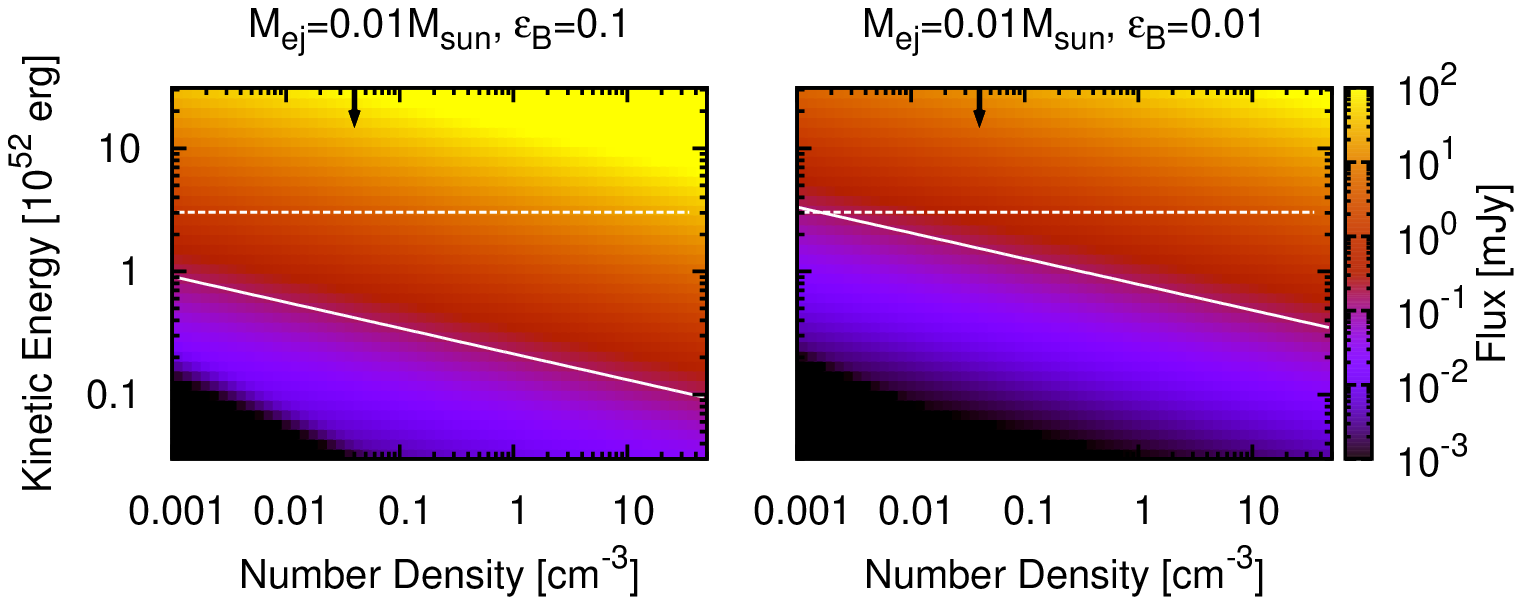}
\includegraphics[width=0.8\textwidth]{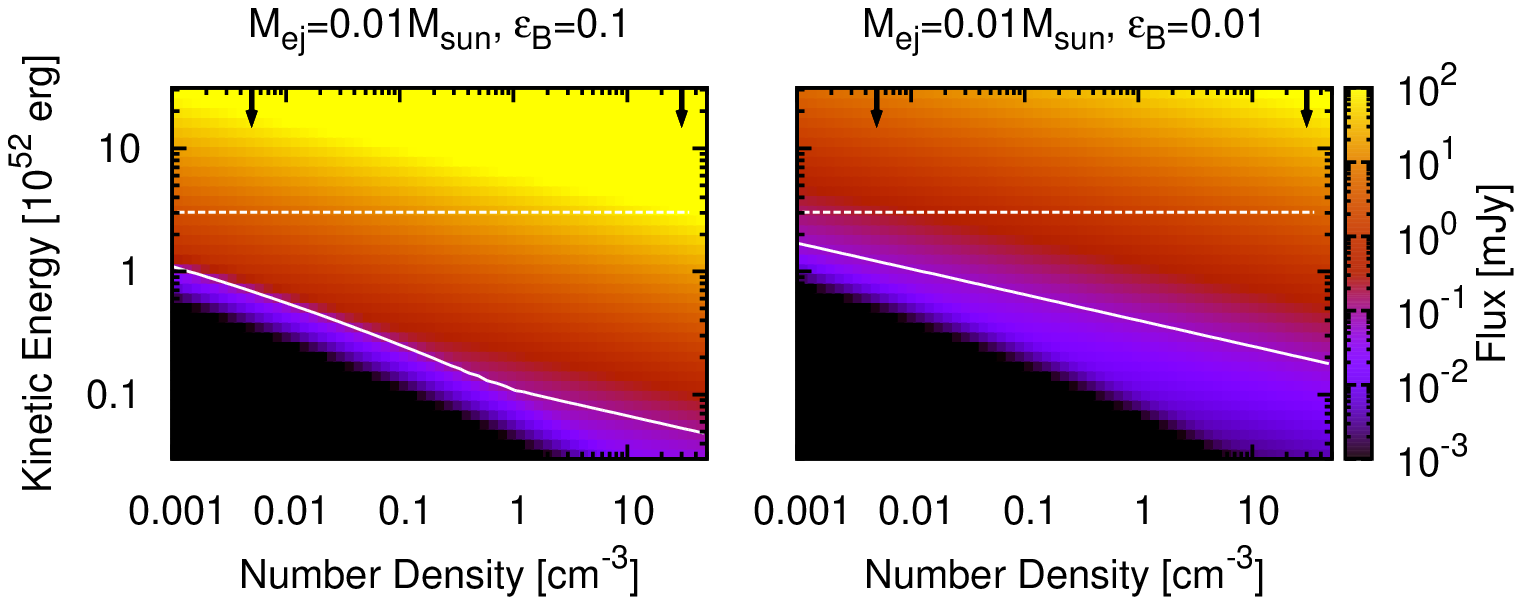}
\caption{Predictions of the radio flux from a macronova (including the
  magnetar model). The
  predictions are for the macronova candidates GRB\,060614 (top
  panel) and GRB\,130603B (bottom panel) at the times we performed our
radio observations (see \S  2 for details). The radio flux of each
event is calculated for a different combination of the kinetic energy
and ISM density. We assume here the fiducial value of the ejecta mass
in the magnetar model of $M_{\rm ej}=0.01\,{\rm M_{\odot}}$. We also
assume $\epsilon_{\rm e}=0.1$ and use both $\epsilon_{\rm B}=0.1$
(left panel) and $\epsilon_{\rm B}=0.01$ (right panel). The dashed
white line represent the fiducial value of the kinetic energy in the
magnetar model, $E_{k}=3\times 10^{52}$\,erg. The solid white lines
represent our observational limits. The arrows represent the ISM
density value (or value range) measured based on the observed afterglow
properties (see \S4).}
\label{fig:phase_space}
\end{figure*}

Next, we calculate (following  Hotokezaka \& Piran 2015)  the expected radio flare signature, using a range of values for $E$, and $n$. We repeat the calculations using
two distinct values for the microphysical parameter, $\epsilon_{\rm B}$,
i.e., $\epsilon_{\rm B}=0.1$ and $\epsilon_{\rm B}=0.01$. The expected
radio signal was calculated for each of the macronova candidates, separately, at the frequencies in which they were
observed. Our
predictions of the radio signals are presented in
Figure~\ref{fig:phase_space}. 

\begin{figure*}[!ht]
\centering
\vspace{20pt}%
\includegraphics[width=0.8\textwidth]{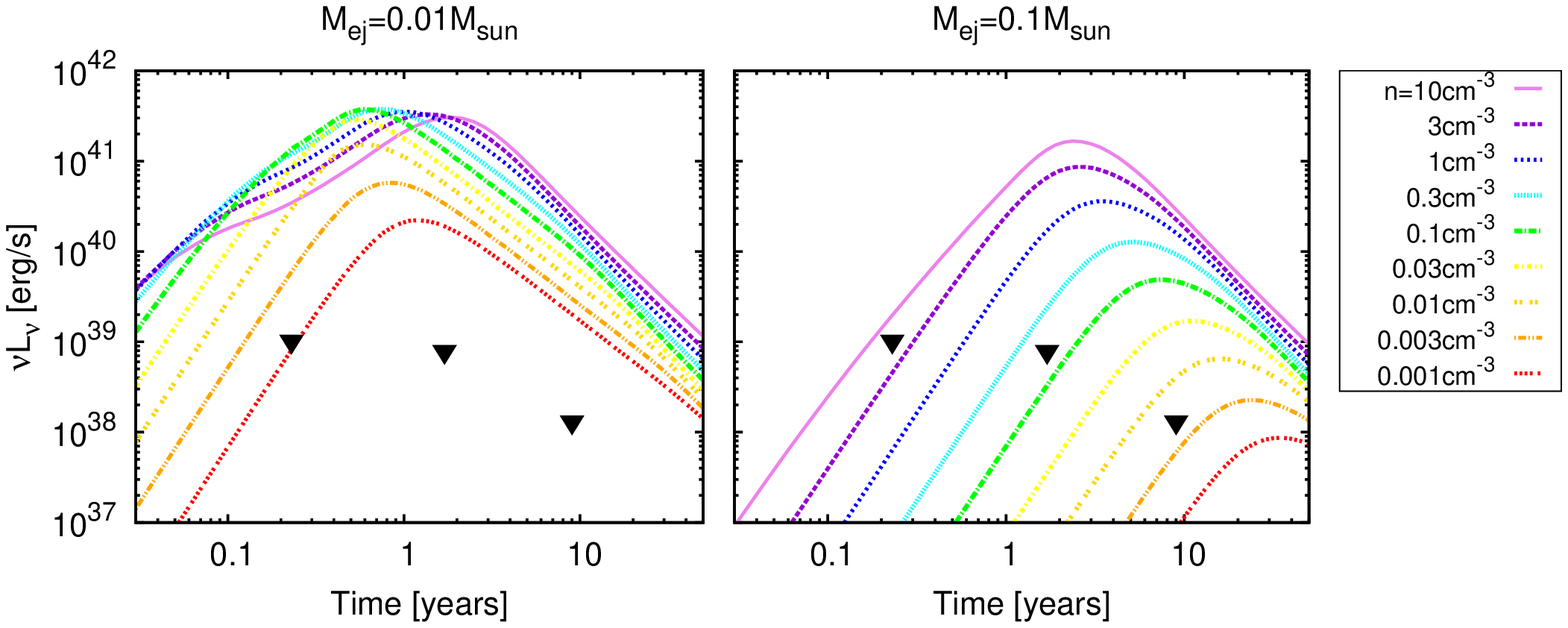}
\caption{Predicted radio light curves in the magnetar scenario when
  both relativistic effects and synchrotron self-absorption are
  included. The light curves are calculated assuming kinetic
  energy of $E_{k}=3\times 10^{52}$\,erg and a range of ISM densities
(see legend). Solid triangles represent the late-time radio
observations (see \S~\ref{sec:obs}).}
\label{fig:lc}
\end{figure*}

Figure~\ref{fig:lc} shows examples of radio light curves specifically for the
fiducial magnetar model with an energy of $E=3\times
10^{52}$\,erg and ejecta mass of $M_{\rm ej}=0.01, ~0.1\,{\rm M}_{\odot}$ for various ISM density values. The relativistic effects alone 
shorten the peak time, compared to the Newtonian case, by a factor of
$\sim 20$ (for $\Gamma = 3$). Adding the effect of synchrotron-self absorption, however, prolongs the peak time. Thus, the combined effect of
synchrotron self absorption and relativistic motion on the peak time
is only a factor of a few, compared to the Newtonian case. The
peak flux can also vary by an order of magnitude. If we take the
case of $n=0.1$ as an example, the peak luminosity and time in the
naive Newtonian case would have been $\approx 2\times 10^{40}$\,erg/s, and
$\approx 930$\,days, compared to $\approx 4\times 10^{41}$\,erg/s, and
$\approx 200$\,days, in the full relativistic calculation.

\section{Comparison of the magnetar model with observations}
\label{sec:comparison}

As seen in Figure~\ref{fig:phase_space}, we can rule out a large
fraction of the $E - n$ phase space for both GRB\,130603B and
GRB\,060614. A main uncertainty in the determination of radio flare signals involves
the external density of the ISM. The surrounding circumburst density
is typically determined from analysis of the GRB's afterglow. However, this determination 
typically suffers from numerous uncertainties and degeneracies between this density 
and other afterglow parameters (in particular with $\epsilon_{\rm B}
$). For example, Fong et al. (2014)  analyzed X-ray, optical and radio observations of
the afterglow of GRB\,130603B. They find that the possible circumburst density ranges from $0.005$ to $30\,{\rm cm}^{-3}$. This large range
of uncertainty demonstrates the difficulty in estimating the density even when afterglow information is available in three bands. 
Xu et al. (2009) have analyzed the afterglow of GRB\,060614. They find
that a density of $0.04\,{\rm cm}^{-3}$ is consistent with the data
but they do not try to 
bracket it. The range of values of the ISM densities for both
GRB\,130603B and GRB\,060614 are within the range that we have
discussed here, and are both sufficiently large to rule out the
canonical magnetar model.

\begin{figure*}[!ht]
\centering
\includegraphics[width=0.7\textwidth]{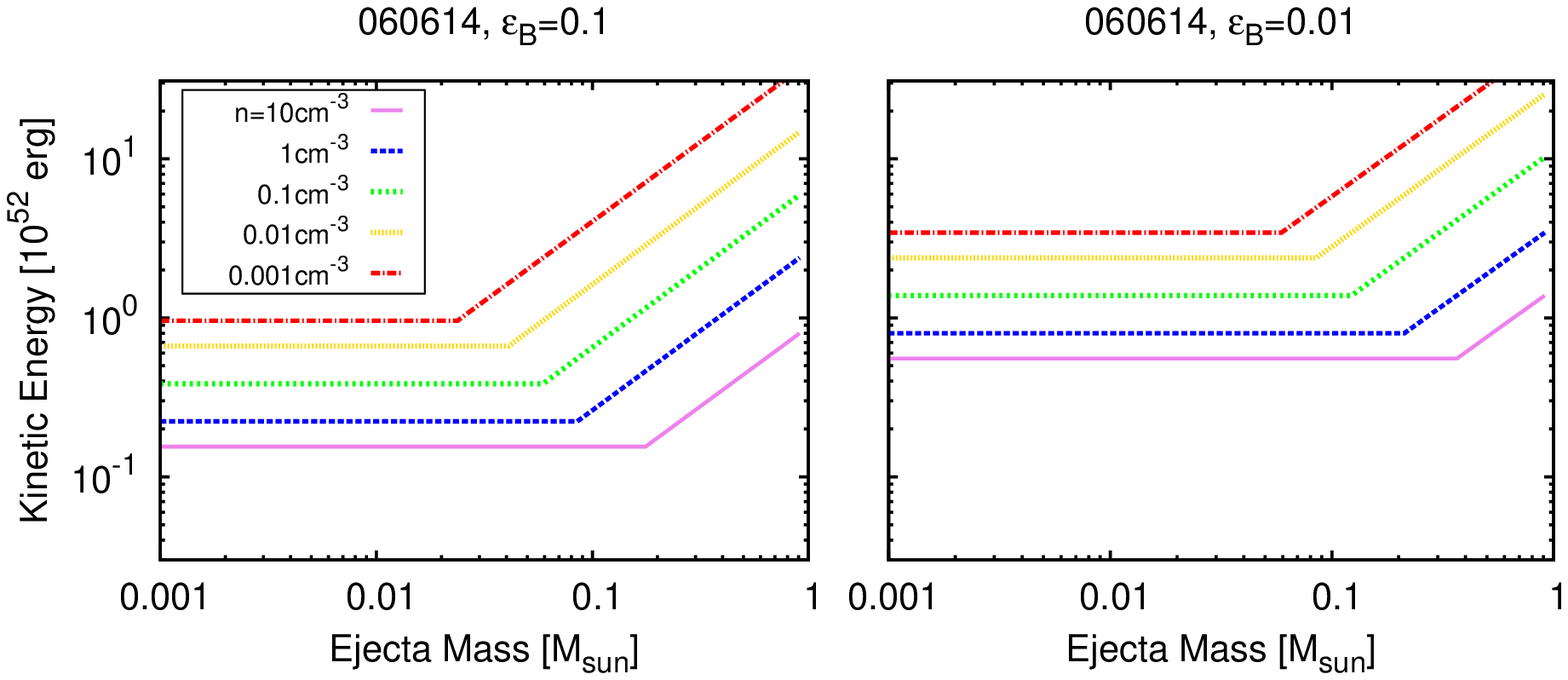}
\includegraphics[width=0.7\textwidth]{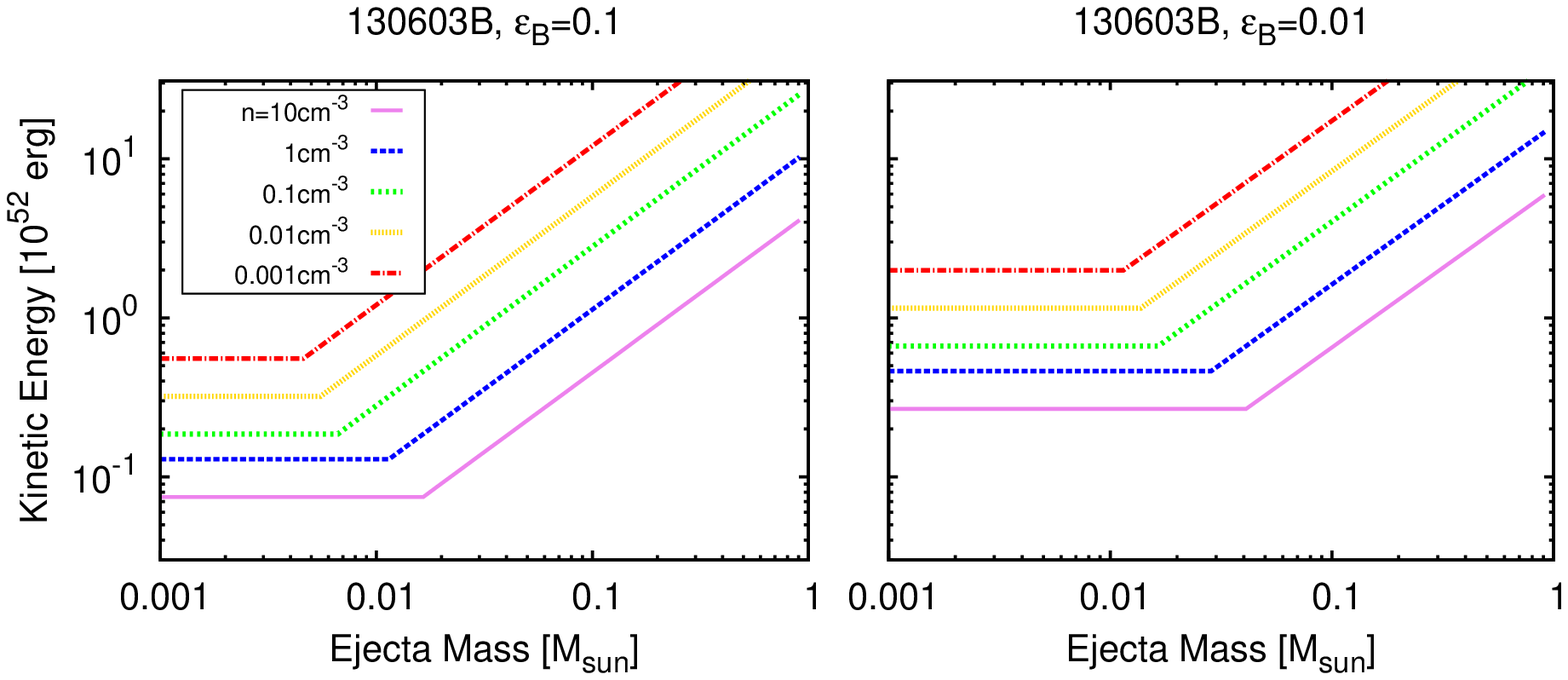}
\caption{The allowed phase space of energy and ejecta mass  (including
  the magnetar and non-magnetar scenario) of the two macronovae
  candidates GRB\,060614 (top
  panel) and GRB\,130603B (bottom panel). The allowed phase space for
  different ISM density values (different line colors) is below each
  density line. The allowed phase space is based on our late-time
  radio observations. We also
assume $\epsilon_{\rm e}=0.1$ and use both $\epsilon_{\rm B}=0.1$
(left panel) and $\epsilon_{\rm B}=0.01$ (right panel).}
\label{fig:full_space}
\end{figure*}

In light of the uncertainty in the ISM density and the microphysical
parameters, we present in Figure~\ref{fig:full_space} different
areas in the ${\rm M}_{\rm ej}$ - $E$ phase space that can be ruled
out for various ISM density and $\epsilon_{\rm B}$ values. This large
phase space, as in Figure~\ref{fig:phase_space}, accounts not only for
the magnetar scenario (discussed below) but also for  
the cases where there is no additional energy injection such as the ``standard"  non-relativistic macronova
scenario presented in Nakar \& Piran (2011).

Assuming that a magnetar output energy is $3\times 10^{52}$\,erg, 
then even for a very low ISM density
$n=0.001$\,cm$^{-3}$ and for a relatively low energy conversion of
shockwave 
energy to magnetic fields, $\epsilon_{B}\sim 0.01$, the expected radio
signal at the time of our radio observations for both events are above
our detection limits. Given that we did not detect any radio emission,
this rules out the fiducial magnetar model for macronova
events associated with GRBs.

It is worth mentioning that the above conclusion is based on
the assumption of spherical symmetry.  Deviations from spherical symmetry, that are expected, would
reduce somewhat the signal and delay the peak time (Margalit \& Piran,
2015). However this amounts only to about $10\%$ difference in peak
luminosity and a factor of $\sim 2$ in peak time. 
We cannot rule out a magnetar with a
large mass ejection ($> 0.1\,{\rm M}_{\odot}$), in low density
environment, by the absence of radio emission. The velocity of this
large ejecta mass will be non-relativistic and is expected to produce
weak emission below our detection limits (Figure~\ref{fig:lc}). 
Other cases where the radio emission can be highly
suppressed is an even more extreme case, where a minute amount of
energy is converted in the shock to magnetic fields, i.e.,
$\epsilon_{B} \ll 
0.001$. Atypical high ISM density will also lead to a suppression of the
radio signal as the optical depth will increase.

\section{Summary}
\label{sec:sumary}

Compact binary mergers are expected to be followed by a macronova
emission and long-lasting radio emission. 
In this paper we have searched for this radio signal including the one
which is predicted specifically by the magnetar scenario. In this latter case, a merger
results in highly magnetized ns that deposits energy into a small
amount of ejecta mass that  becomes relativistic. If this
relativistic ejecta interacts with an ISM that is not too dilute, it is expected
to produce a bright radio emission which will peak over time scales of
months to years. 

Our search was focused on two GRBs
(GRB\,130603B \& GRB\,060614) that were the first to exhibit a
macronova-like emission, thus indicating 
the ejection of a small amount of mass, a condition needed for the
late production of a radio flare. Therefore, we have observed these GRB
positions at late times with the VLA  and the ATCA telescopes. 
Our radio observations resulted in null-detections. Comparing the
predicted radio emission with our upper
limits, we can rule out a wide range of kinetic energies,
ejecta masses, ISM densities and microphysical parameters. As shown in
Figures~1, 2, and 3, the
range of parameters we rule out includes the canonical magnetar
model.

A previous search for magnetar radio emission from sGRBs has been
performed by Metzger \& Bower (2013). They observed $7$ sGRBs
within $1 - 3$ years after discovery with the VLA but did not detect
any emission. They  used their non-detections to constrain
the merger magnetar scenario as well. However, their work is different
from ours in several ways. First, they have used the Newtonian calculations following Nakar
\& Piran (2011) but with $\beta=0.8$ and kinetic energy of $3\times 10^{52}$\,erg/s. Thus, they have not accounted for
relativistic effects and did not explore a wide range of ejecta
masses. Given these limitations and the lower observational
sensitivities (due to the old capabilities of the VLA), Metzger \&
Bower (2013) only ruled out magnetar scenarios with densities
above $n=0.03\,{\rm cm^{-3}}$. 
Our observed sample is also
different since the sGRBs that we observed have been associated with
macronova emission, previously not observed in other sGRBs.

As discussed above, our conclusion is limited by several
factors. While, we use a wide range of values for the model
parameters, there are still extreme parameters under which the magnetar
model is consistent with our observations. This includes, extremely high (or
low) ISM density, extremely low values ($<0.001$) of $\epsilon_{\rm
  B}$, and extremely small ejecta mass. 
Given these limitations and the fact that we studied
only two macronova events, provides further motivation to undertake a
large campaign of carefully designed late-time radio observations of sGRBs. 

\pagebreak
\section*{Acknowledgments}
\label{sec:ACK}

We thank the ATCA and the VLA staff for promptly scheduling
the observation of these targets of opportunity. 
The National Radio Astronomy Observatory is a facility of the National Science Foundation operated under cooperative agreement by Associated Universities, Inc.
Research leading to these
results has received funding from the EU/FP7 via ERC grant 307260; ISF, Minerva,
and Weizmann-UK grants; as well as the I-Core Program of the Planning and Budgeting Committee
and the Israel Science Foundation, and the Israel Space Agency. Parts of this research were conducted by the Australian Research Council Centre of Excellence for All-sky Astrophysics (CAASTRO), through project number CE110001020.

\end{document}